\documentclass[]{article}

\usepackage{arxiv}

\usepackage[utf8]{inputenc} % allow utf-8 input
\usepackage[T1]{fontenc}    % use 8-bit T1 fonts
\usepackage{hyperref}       % hyperlinks
\usepackage{url}            % simple URL typesetting
\usepackage{booktabs}       % professional-quality tables
\usepackage{amsfonts}       % blackboard math symbols
\usepackage{nicefrac}       % compact symbols for 1/2, etc.
\usepackage{microtype}      % microtypography
\usepackage{graphicx}
\usepackage{natbib}
\usepackage{doi}
\usepackage{arydshln}
\usepackage{amsmath}
\usepackage{subcaption}
\usepackage{multirow}
\usepackage{authblk}
\usepackage{bm}

\title{Is 1:1 Always Most Powerful? \\ Why Careful Determination of Allocation Ratios Matters in Trial Design
}

\author[1,2,3]{Lukas Pin\textsuperscript{*}}
\author[1,2,3]{Stef Baas}
\author[1,2]{David S. Robertson}
\author[1,2,3]{Sofía S. Villar}

\affil[1]{Efficient Study Design Group, MRC Biostatistics Unit, University of Cambridge, Cambridge, UK, \url{www.mrc-bsu.cam.ac.uk/research/efficient-study-design}}
\affil[2]{Adaptive Designs Working Group (ADWG), MRC-NIHR Trials Methodology Research Partnership (TMRP), \url{www.methodologyhubs.mrc.ac.uk/about/working-groups/adaptive-working-group/}}
\affil[3]{Randomization Working Group, \url{www.randomization-wg.org}}
\affil[*]{\textit{Corresponding author:} \href{mailto:lukas.pin@mrc-bsu.cam.ac.uk}{lukas.pin@mrc-bsu.cam.ac.uk}}

\renewcommand{\shorttitle}{\textit{Why Careful Determination of Allocation Ratios Matters in Trial Design}}

\hypersetup{
pdftitle={Unequal Allocation Merits Broader Consideration},
pdfsubject={biostatistics, RAR},
pdfauthor={Lukas Pin, Stef Baas, David Robertson, Sofia Villar},
pdfkeywords={Wald test, Patient-benefit, Fixed Randomisation, Unequal Randomisation Ratios, Neyman Allocation, Expected Patient Outcomes},
}

\begin{document}
\maketitle

\begin{abstract}
The principle of allocating an equal number of patients to each arm in a randomized controlled trial remains widely believed to be optimal for maximising statistical power. However, this long-held belief only holds true if the treatment groups have equal outcome variances, a condition that is often not met or, is simply not assessed in practice. 
This paper reasserts the fact that a departure from a 1:1 ratio can maintain or improve statistical power while increasing the benefits to participants. The benefit is particularly self-evident for binary and time-to-event endpoints, where variances are determined by the assumed success or event rates.
To illustrate this, we present two case studies: a small-scale metastatic melanoma trial with a binary endpoint and a larger trial evaluating virtual reality for pain reduction with a continuous endpoint. Our simulations compare equal randomisation, preplanned fixed unequal randomisation, and response-adaptive randomisation targeting Neyman allocation. Results show that unequal allocation can increase the proportion of patients receiving the superior treatment without reducing power, with modest power gains observed in both binary and continuous settings, highlighting the practical relevance of optimised allocation strategies across trial types and sizes.
\end{abstract}


\section{Introduction}

Many trialists and statisticians still strongly adhere to the default assumption that equal allocation (a 1:1 ratio) maximizes statistical power, often neglecting to determine the appropriate allocation ratio based on trial specifics. Many review articles frequently recommend the 1:1 ratio by convention \citep{Dumville2006, DIBAODINA20141070, Berger2021, freidlin2024two}, for instance:
\begin{quote}
``balanced group sizes will maximise a study’s statistical power" \citep{Dumville2006}.
\end{quote}
The principle is further reinforced by influential textbooks \citep{Pocock2013, Friedman2015}. 
For example, Chapter 9 of \cite{Pocock2013} details standard power calculations where formulas implicitly assume a 1:1 patient allocation, yielding $n_C=n_T$ patients per treatment arm (control ($C$) and treatment ($T$)). Building the equal allocation ratio into standard sample size formulae calculations, whether intentional or not, contributes to elevating the 1:1 ratio as the default, on top of making it inherently less straightforward to compute optimal sample sizes per arm. 
Together, these factors  subtly discourage statisticians from calculating the true optimal allocation, i.e.\ the mathematically most powerful choice. In practice, there is evidence that the 1:1 ratio is by far the most common ratio used. For example, \cite{Gupta2021} found that, among all two-arm oncology trials registered on ClinicalTrials.gov between 2010 and 2019 (inclusive), 74\% employed a 1:1 allocation ratio.

The rationale for determining an allocation ratio is multifaceted \citep{Avins1998}, reflecting not just statistical considerations but also ethical, financial, and practical ones, including patient recruitment. However, many of these arguments are implicitly shaped by the assumption that equal allocation always maximizes statistical power \citep[e.g.][]{torgerson1997unequal, wu2014du-random, NAY2024107484} —a claim that does not hold in general. 
%These arguments are often conflated with the ethical and practical justifications being reinforced by the assumption that equal allocation is always the most statistically powerful design:
\begin{quote}
    ``Trials using unequal allocation will therefore either have less statistical power or will be more expensive and entail exposing more patients than necessary to a novel intervention and research procedures." 
    \citep{Hey2014}.
\end{quote}

This long-held assumption that equal allocation maximises power is conditional. Its mathematical validity rests on the crucial prerequisite that the outcome variances within the compared groups are equal. However, this assumption of homoscedasticity is often questionable in practice and may naturally be violated in many settings \citep{Ruxton2006}. When this condition of equal variances is not met, the optimality of equal allocation is no longer guaranteed. 
For commonly used primary endpoints like binary (yes/no) and time-to-event, the inherent spread of the outcomes (variance) is mathematically determined by the expected success or event rates. Therefore, the anticipated treatment difference does more than just define the minimal clinically important difference (MCID); it mathematically implies that the variance of the control and experimental arms will necessarily be different, which then enables the direct calculation of an optimal unequal allocation ratio. Consequently, the standard 1:1 allocation is almost always automatically suboptimal for these endpoints.
This means that for such endpoints, by moving away from a 1:1 ratio, we may achieve superior performance in terms of power, often alongside other benefits such as patient outcomes. This superior performance can be achieved through allocation strategies such as fixed unequal allocation (by sampling more from the more variable arm) or response-adaptive randomisation (RAR), where allocation probabilities are dynamically adjusted as data accrue and variance estimates improve.

The objective of this paper is to challenge the conventional default application of equal allocation and convey a fundamental point: that equal allocation is not a universal optimum. We argue that a well-reasoned allocation ratio, informed by beliefs about outcome variances (even if it deviates from equal allocation), is a superior option to defaulting to 1:1 without careful consideration of outcome group variability. This conclusion is particularly relevant for confirmatory clinical trials using binary or time-to-event endpoints, as these remain among the most frequently used measures in current statistical practice, for example, \cite{thompson2025RR} reports that 28\% of confirmatory two-arm superiority clinical trials employ a binary endpoint. The principle also holds true for continuous endpoints where the variance may reasonably differ between the treatment and control arms.

\section{Determining the Optimal Allocation Ratio for Maximizing Power}\label{sec-meth}

The allocation ratio that maximises statistical power for a given test statistic depends on:
\begin{enumerate}
    \item the type of endpoint (binary, discrete, continuous),
    \item the measure of interest (e.g., mean difference, odds ratio),
    \item and, most critically, the true variance of the outcome within each treatment arm.
\end{enumerate}

We illustrate how the above elements interplay when aiming to maximise power using the conventional 
%and widely-used 
Wald test for comparing two group means.
For this test, it has been shown that comparing groups under heteroscedasticity, sampling with an unequal ratio can increase statistical power \citep{Tschuprow1923, Neyman1934, Robbins1952, Rosenberger2001}. By treating the question as a formal optimization problem, one can derive the optimal allocation strategy for the Wald test. This principle, known as \textit{Neyman allocation}, dictates that to maximize power for a fixed total sample size, the allocation ratio between the treatment and control arms should be proportional to the standard deviations of the outcomes in each group:
\begin{equation}\label{eq:NeymannAll}
    \frac{n_T}{n_C} = \frac{\sigma_T}{\sigma_C}
\end{equation}
It follows from Formula \eqref{eq:NeymannAll} that ${n_T}={n_C}$ is only optimal when the true standard deviations are equal ($\sigma_T = \sigma_C$). Otherwise, maximising statistical power requires allocating more patients to the arm with higher expected variance.

\begin{figure}[!htbp]
    \centering
    \includegraphics[width=1\linewidth]{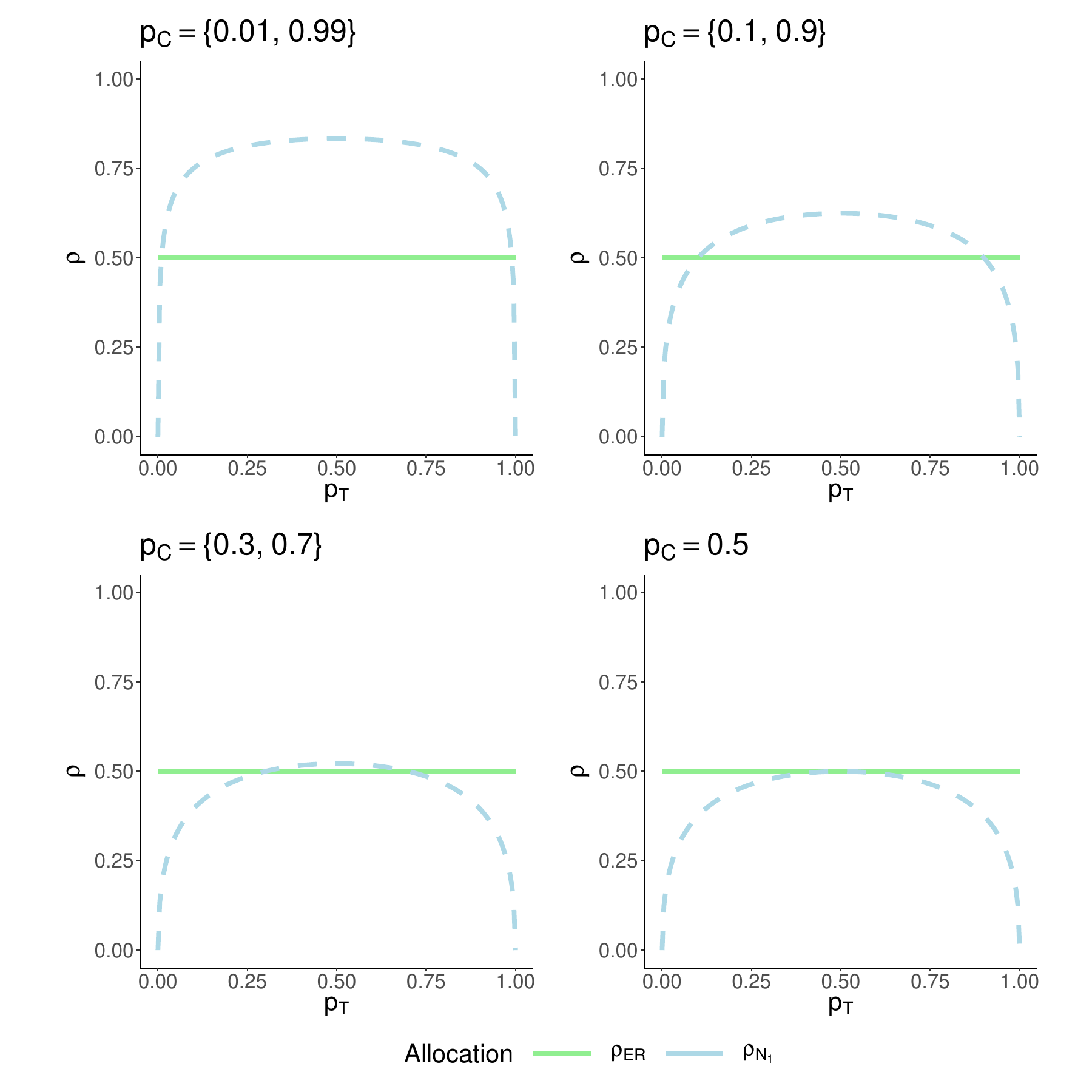}
    \caption{The horizontal axis shows the varying value of $p_T$ while $p_C$ takes a fixed value in each sub figure. For each combination of $p_T$ and $p_C$ the vertical axis shows the value of the theoretical Neyman allocation ($\rho_{N_1}$) and equal randomisation ($\rho_{\text{ER}}$). Notice that as Neyman allocation is symmetric around 0.5 each sub figure represent two $p_C$ values.}
    \label{fig:binaryNeyman}
\end{figure}

In the case of binary endpoints, the variance $\sigma_i^2 = p_i(1-p_i)$ within each arm ($i=C,T$) is determined by its success rate, $p_i$. Applying the principle of Neyman allocation reveals a critical nuance: ER can only be optimal under  homoscedasticity. This key assumption only holds true for the null hypothesis ($p_C=p_T$) or in those points of the parameter space where the success rates are symmetric around 0.5 (i.e., $p_C = 1 - p_T$), as we illustrate in Figure \ref{fig:binaryNeyman}.
Figure \ref{fig:binaryNeyman} illustrates the mathematical tension between optimal design and conventional practice. The curved line represents the optimal allocation proportion necessary to maximize statistical power for a binary endpoint under a given parameter configuration, plotted as a function of the treatment success rate $p_T$ (taking values within $[0,1]$) while holding the control rate $p_C$ fixed. For comparison, the flat line shows the universally applied 1:1 ratio. The most striking observation from Figure \ref{fig:binaryNeyman} is the substantial lack of alignment between the theoretically optimal allocation and the 1:1 default. They only intersect at the few specific points where homoscedasticity mathematically holds true. The rarity of these intersection points demonstrates that adhering to the 1:1 allocation ratio is suboptimal for power in the vast majority of two-arm trials with binary outcomes.
This figure also elucidates the possibility of a conflicting situation where maximising statistical power requires the allocation ratio to shift in a direction that favours the currently worse-performing treatment arm (i.e., in cases where the most variable arm is also the worst one). 

This highlights a broader principle: when maximising statistical power, the optimal allocation depends heavily on the context and therefore merits careful consideration beyond simple default practices. For instance, with time-to-event data that is exponentially distributed, and where a longer time is a positive outcome, Neyman allocation always assigns more patients to the superior arm \citep{Antognini2020}. Furthermore, the choice of effect measure and the statistical test itself alter the optimal ratio \citep{Pin_Deming2024}. 

More generally, power is optimized by the allocation that maximizes the test statistic's non-centrality parameter \citep{hu2003optimality}, which for the Wald test leads to Neyman allocation \citep{Rosenberger2001}.
We conclude this section with the theoretical insight that equal allocation can only be expected to be optimal in terms of statistically power only under homoscedasticity while homoscedasticity is itself a rare possibility for some commonly used outcome types; its practical implications will be investigated in the next section.

\section{Case studies: From Optimal Statistical Power to Informed Allocation}\label{Sec:practicalExamples}

The preceding discussion established the fundamental point that an equal allocation ratio is mathematically optimal only under rare conditions for key clinical trial endpoints. However, moving from theory to practice requires addressing a critical question: how much does this suboptimality truly cost, and how sensitive is statistical power amongst other metrics to deviations from the calculated optimal ratio? To answer this, we present  simulations where we compare the statistical power attained with a Wald test for 3 cases: equal randomisation (ER); a pre-planned fixed unequal randomisation (FUR) and RAR targeting Neyman allocation. 
Since our simulated outcomes are independent and identically distributed, our evaluation of ER and FUR is unaffected by the choice of randomisation procedure, so long as the final allocation count is achieved. Our findings therefore apply to any method that achieves the target, including the truncated binomial design, big stick design, permuted block design, and the random allocation rule \citep{Berger2021}.

\subsection{Case Study with binary endpoint: metastatic melanoma trial}\label{Sec:practicalExamples1}

We illustrate the difference between ER and FUR using the trial by \citet{Chapman2011}, which compared vemurafenib against dacarbazine in patients with metastatic melanoma harboring the \textit{BRAF V600E} mutation. Metastatic melanoma is an aggressive cancer with historically poor prognosis, and the observed response rate in the trial to dacarbazine—the standard chemotherapy at the time—were around 5\%. In contrast, vemurafenib, a targeted BRAF inhibitor, produced a markedly higher observed response rate of approximately 48\%.  

For our re-analysis, we consider a simplified binary endpoint representing tumor response (complete/partial response or stable disease = 1, progression = 0). Because the original trial enrolled hundreds of patients, achieving an incredibly but also exceptionally large power to detect such a treatment effect, we reran the design with a reduced total sample size of 36 patients, chosen to achieve roughly 90\% power using a Wald test in an ER design, while keeping the response rates the same as those observed. While real-world confirmatory trials often require a larger sample size to address crucial secondary questions (such as safety), we deliberately use a smaller sample size in this simulation to isolate and illustrate the impact of the optimal allocation ratio on the primary endpoint's statistical power.

As before, the primary analysis is based on a Wald test at a two-sided 5\% significance level. The FUR design assigns patients at a 1:2 ratio between control and treatment, approximating the Neyman allocation based on the standard deviation ratio.  

\begin{table}[!htbp]
\centering
\caption{Comparison of allocation strategies for the binary endpoint trial with $n=36$. Metrics include type-I error rate ($p_C=p_T=0.05$), power ($p_C=0.05$, $p_T=0.48$), the proportion of patients on the superior arm ($n_1/n$), and the Expected Number of Successes (ENS) under the alternative. The nominal significance level is $\alpha = 5\%$. Values in parentheses indicate deviations from $\alpha$ (for type-I error rate) and from ER (for power). The Monte-Carlo error for type-I error rate and power is less than $0.2\%$ \citep{Morris2019}.}
\label{tab:binarytrial}
\vspace{0.2em}
\begin{tabular}{@{}lcccc@{}}
\toprule
\textbf{Design} & \textbf{Type-I Error Rate} & \textbf{Power} & \textbf{$\bm{n_1/n}$} & \textbf{ENS} \\
\midrule
ER        & 0.8\% {\small(-4.2)} & 91.7\% {\small(0)}  & 0.50 & 9.5 \\
FUR (1:2) & 2.3\% {\small(-2.7)} & 93.7\% {\small(+2.0)} & 0.67 & 12.1 \\
RAR       & 8.3\% {\small(+3.3)} & 95.3\% {\small(+3.6)} & 0.80 & 14.2 \\
\bottomrule
\end{tabular}
\end{table}

As shown in Table \ref{tab:binarytrial}, moving away from equal allocation increases statistical power and assigns more patients to the superior treatment arm. Using FUR improves power by $2$ percentage points over ER and yields $2.4$ more expected successes, which is substantial in such a small trial with a severe endpoint. 

Testing in small samples with a binary endpoint introduces challenges due to discreteness and the breakdown of asymptotic assumptions, compromising the quality of the Central Limit Theorem approximation. Table \ref{tab:binarytrial} shows how the Wald test in this setting is overly conservative both for ER and FUR.
It is worth noting that while exact tests like Fisher's or the optimized Berger-Exner (Boschloo) methods avoid reliance on these approximations, they often result in overly conservative inference (all yield an actual $\alpha$ of $\leq$1\% with power <90\% in this case). In view of this, the conservatism of the standard Wald test for ER appears to be beneficially mitigated by the use of FUR, while still remaining under the target $\alpha$-level.

Pinpointing the optimal unequal ratio for FUR at the trial's outset is obviously challenging when the endpoint is not binary or time-to-event. This hurdle can be overcome by using RAR instead. This may also help if there is uncertainty on the baseline success or event rate. 
In our case study we illustrate how allowing for estimating the optimal ratio from the trial data, using the efficient randomized-adaptive design of \citet{Hu2009} as an RAR design to target Neyman allocation increases power by 4 percentage points.
However, using this type of RAR for binary endpoints in combination with the Wald test can lead to serious type-I error rate inflation. Methods exist to control the type-I error rate when using RAR \citep{pin2025revisitingoptimalallocationsbinary, baas2025computationalmethodtypei}. In this case study the RAR design incorporates a ``burn-in" period where the first 8 patients are allocated equally, after which the randomization probabilities are adapted.

\bigskip 

\subsection{Case Study with continuous endpoint: virtual reality for pain reduction during outpatient hysteroscopy}\label{Sec:practicalExamples2}

We next illustrate the performance of ER, FUR, and RAR in a setting with a continuous outcome, using data inspired by the randomized controlled trial by \citet{jcm12113645}. The original trial investigated whether the use of virtual reality (VR) immersion could reduce pain during outpatient hysteroscopy compared to standard care. Pain intensity was measured on a visual analogue scale (VAS) ranging from 0 (no pain) to 10 (worst imaginable pain). A statistically and clinically significant reduction in mean pain scores of approximately 1.5 points in the VR group compared to the control group was reported.  

Because the original study was highly powered due to the large treatment effect, we redesigned the trial for our simulation study to make the comparison between allocation strategies more meaningful. Specifically, we reduced the assumed treatment difference to a more moderate 0.5 points and kept the total sample size at $n=154$ participants (77 per group under ER). As before, we use a two-sided Wald test at a 5\% significance level.

To align with our framework where higher outcomes correspond to better responses, we multiply the outcomes by $-1$ so that a lower pain score (more effective treatment) corresponds to a higher mean response value in our simulations. Thus, under the null hypothesis, both arms have a mean of $\mu_C=\mu_T=-3.972$, and under the alternative, the treatment arm has a mean of $\mu_T=-3.472$. The outcome variance is estimated from the reported standard deviations in the original study.

The FUR design assigns patients at a 1:2 ratio between control and treatment arms, reflecting the approximate Neyman allocation derived from the standard deviation ratio. For the RAR design, we used the efficient randomized-adaptive procedure of \citet{Hu2009} targeting Neyman allocation, including a burn-in period of 15 patients per arm before adaptation begins. 

\begin{table}[!htbp]
\centering
\caption{Comparison of allocation strategies for the continuous outcome trial. Metrics include type-I error rate ($\mu_C=\mu_T=-3.972$), power ($\mu_C=-3.972$, $\mu_T=-3.472$), the proportion of patients on the superior arm ($n_1/n$), and the Expected Mean Response (EMR) under the alternative. The nominal significance level is $\alpha = 5\%$. Values in parentheses indicate deviations from $\alpha$ (for type-I error rate) and from ER (for power). Monte-Carlo error for type-I error rate and power is less than $0.2\%$ \citep{Morris2019}.}
\label{tab:continuoustrial}
\vspace{0.2em}
\begin{tabular}{@{}lcccc@{}}
\toprule
\textbf{Design} & \textbf{Type-I Error Rate} & \textbf{Power} & \textbf{$\bm{n_1/n}$} & \textbf{EMR} \\
\midrule
ER        & 5.1\% {\small(+0.1)} & 92.0\% {\small(0)}   & 0.50 & $-3.712$ \\
FUR (1:2) & 5.3\% {\small(+0.3)} & 95.6\% {\small(+3.6)} & 0.67 & $-3.638$ \\
RAR       & 5.4\% {\small(+0.4)} & 95.6\% {\small(+3.6)} & 0.69 & $-3.627$ \\
\bottomrule
\end{tabular}
\end{table}

As shown in Table \ref{tab:continuoustrial}, both FUR and RAR modestly improve power compared to ER, with gains of approximately 3.5 percentage points. The unequal designs also yield slightly better expected mean responses, as a higher proportion of participants receive the superior treatment. In this example, RAR performs similarly to the pre-specified FUR, as the standard deviation ratio is stable and well estimated, yet it retains the flexibility to adjust allocation dynamically should observed variability differ from expectations.

\bigskip

\addtolength{\textheight}{-.2in}%

\section{Discussion}\label{Sec:Discussion}

Whether and, crucially, by how much equal allocation can be superior in terms statistical power for a given statistical test depends on several key factors:
\begin{enumerate}
    \item the endpoint type, 
    \item the total sample size,
    \item the magnitude of the treatment effect difference,
    \item the relationship between the variances of the treatment arms.
\end{enumerate}
While Neyman allocation might indicate a departure from a 1:1 ratio is optimal, the benefits in terms of statistical power become less pronounced when a study is already robustly powered with ER. This may be the case for studies with a large sample size or an expected large treatment effect. Our key objective was to demonstrate that implementing an unequal allocation ratio—often highly desirable for patient benefit —is not inherently detrimental to statistical power.
We found that a substantial reduction in power, relative to equal randomization, occurs only when the chosen allocation ratio deviates markedly from Neyman allocation, and not in an already overpowered study. In particular, power losses are observed only when a sub-optimal ratio is skewed in the wrong direction or excessively unbalanced—for example, a 2:1 allocation (yielding 85.4\% power) or a 1:9 allocation (80.8\% power) in our first case study—compared with 91.7\% power under equal randomization (see Table \ref{tab:binarytrial}).

We found that a substantial reduction in power (relative to equal randomization) occurs only when the chosen allocation ratio substantially misses the optimal (Neyman) proportion and not in an overpowered study. Specifically, choosing a sub-optimal ratio that is skewed in the wrong direction or excessively skewed, such as a 2:1 ratio (resulting in $85.4\%$ power) or 1:9 ratio ($80.8\%$ power) in our first case study, is required to see significant penalties compared to 91.7\% power under ER, see Table \ref{tab:binarytrial}.

%, e.g. in Case Study 2 choosing a 2:1 ratio ($76.2\%$ power) or 1:19 ratio ($87.3\%$ power).

While power increased in our examples, FUR or RAR aiming for Neyman allocation often yields power comparable to ER, especially when the optimal allocation is near 1:1. For binary endpoints, Figure~\ref{fig:binaryNeyman} shows that optimal Neyman allocation deviates little from equal allocation when success probabilities are close to 0.5. %Importantly, this implies that unequal allocation is rarely detrimental. 
In many regions of the parameter space, power gains are minimal or nonexistent, but it can still favour participants by assigning more to the better-performing treatment with minimal impact on power. Thus, the value of unequal allocation lies not just in maximizing power, but also in enhancing patient outcomes without decreasing statistical power.
It is important to note that our findings are based on a two-arm setting using the parametric Wald test. Choosing an optimal allocation ratio for more specific or complex scenarios and other tests requires further study, \citep[e.g.][]{JACKSON2025108043}.

Beyond statistical power and participant benefit, unequal allocation may also offer financial and practical advantages. First, if an unequal ratio reduces the total sample size needed to achieve a given power, this directly lowers trial costs. Second, when treatment arms differ in per-patient costs, favouring the less expensive arm can further reduce overall expenditure. From a patient perspective, allocating more participants to the better-performing treatment can reduce burden and, more generally, make trial participation more appealing, potentially speeding enrolment \citep{Meurer2012, Tehranisa2014}. Practically, unequal ratios can also help accommodate real-world constraints such as limited drug supply or differential treatment availability across centres. These factors provide compelling advantages for unequal allocation, and these benefits are often ignored due to the misconception that the practice is inherently detrimental to statistical power.

The potential benefit of unequal allocation extends well beyond simple statistical power. Our analysis demonstrates that failing to consider unequal allocation (e.g. Neyman allocation) —even when its power advantage is minimal— often means sacrificing significant gains in patient welfare and potential economic efficiency. Consequently, we advocate for a fundamental shift in practice: trial designs must move beyond the default 1:1 ratio toward the deliberate determination of the optimal allocation ratio based on trial-specific parameters. The allocation ratio should always be an explicit, data-driven decision assessed against the full spectrum of statistical, ethical, and financial benefits available to the study.

\section*{Acknowledgements}
    \begin{itemize}
   % \item The code used for this paper is available with this paper at the \textit{Clinical Trials} website on Sage. The code can be found at the GitHub repository \url{https://github.com/lukaspinpin/Un-Equal-Randomisation}, and contains the R code necessary to replicate the results presented in the paper.
    \item The members of the working groups listed as affiliations are responsible for the content of this paper. This document does not necessarily represent the consensus or official endorsement of all members of these groups.
    \item The authors acknowledge funding and support from the UK Medical Research Council (grants MC UU 00002/15 and MC UU 00040/03 (SSV, DSR, SB), as well as an MRC Biostatistics Unit Core Studentship (LP) and the Cusanuswerk e.V. (LP). SSV is part of PhaseV's advisory board.
    \item The authors acknowledge the use of large language models (LLMs), specifically Gemini (Google), to assist with generating figures and refining grammar and wording in this paper. The LLMs were not used for data analysis, interpretation, or original scientific writing. All content has been carefully reviewed and verified by the authors, who take full responsibility for the integrity and accuracy of the work.
    \item We thank Alex John London and Stephen Senn for their insightful comments on an earlier draft of this paper. 
    \end{itemize}

\bibliographystyle{abbrvnat} % Alternative: unsrtnat
\bibliography{references}

%\printbibliography

 %%% Uncomment this line and comment out the ``thebibliography'' section below to use the external .bib file (using bibtex) .

%%% Uncomment this section and comment out the \bibliography{references} line above to use inline references.
% \begin{thebibliography}{1}

% 	\bibitem{kour2014real}
% 	George Kour and Raid Saabne.
% 	\newblock Real-time segmentation of on-line handwritten arabic script.
% 	\newblock In {\em Frontiers in Handwriting Recognition (ICFHR), 2014 14th
% 			International Conference on}, pages 417--422. IEEE, 2014.

% 	\bibitem{kour2014fast}
% 	George Kour and Raid Saabne.
% 	\newblock Fast classification of handwritten on-line arabic characters.
% 	\newblock In {\em Soft Computing and Pattern Recognition (SoCPaR), 2014 6th
% 			International Conference of}, pages 312--318. IEEE, 2014.

% 	\bibitem{hadash2018estimate}
% 	Guy Hadash, Einat Kermany, Boaz Carmeli, Ofer Lavi, George Kour, and Alon
% 	Jacovi.
% 	\newblock Estimate and replace: A novel approach to integrating deep neural
% 	networks with existing applications.
% 	\newblock {\em arXiv preprint arXiv:1804.09028}, 2018.

% \end{thebibliography}

\end{document}